\documentclass{article}
\usepackage{spconf,amsmath,graphicx}
\usepackage{amsmath,amsfonts,amssymb}
\usepackage{graphicx}
\usepackage[colorlinks=true,
            citecolor=blue,
            linkcolor=black,
            urlcolor=black]{hyperref}
\usepackage{colortbl}
\usepackage[table]{xcolor}
\usepackage{array}
\usepackage{amssymb}
\usepackage{tikz}
\usepackage{calc}
\usepackage[T1]{fontenc}
\usepackage{adjustbox}
\usepackage{graphicx}
\usepackage{subcaption}
\usepackage{float}
\usepackage{graphicx}
\usepackage{subcaption}
\usepackage{multirow}
\usepackage{booktabs} % for better table rules
\usepackage{hyperref}  % Package for hyperlinks
\usepackage{cleveref}  % Enhanced references (optional)
\usepackage{enumitem}
\usepackage{tikz}
\def\checkmark{\tikz\fill[scale=0.4](0,.35) -- (.25,0) -- (1,.7) -- (.25,.15) -- cycle;} 
\def\scalecheck{\resizebox{\widthof{\checkmark}*\ratio{\widthof{x}}{\widthof{\normalsize x}}}{!}{\checkmark}}
\usepackage{enumitem}
\setlist{nosep, leftmargin=14pt}

\usepackage{mwe} % to get dummy images

\title{CT-AGRG: Automated Abnormality-Guided Report Generation from 3D Chest CT Volumes}

\name{Theo Di Piazza$^{1, 2}$ \quad Carole Lazarus$^{3}$ \quad Olivier Nempont$^{3}$ \quad Loic Boussel$^{1, 2}$}
\address{$^{1}$UCBL 1, INSA Lyon, CNRS, Inserm, CREATIS UMR5220, U1294, Villeurbanne, F-69621, France\\
         $^{2}$Department of Radiology, Croix-Rousse Hospital, Hospices Civils de Lyon, Lyon, France\\
         $^{3}$Philips Clinical Informatics, Innovation Paris, France}
\begin{document}
%\ninept
%
\maketitle

\textit{The rapid increase of Computed Tomography examinations have created a need for robust automated analysis techniques in clinical settings to assist radiologists managing their growing workload. Existing methods generate entire reports directly from 3D CT images, without explicitly focusing on observed abnormalities. This unguided approach can result in repetitive content or incomplete reports. We propose a new anomaly-guided report generation model, which first predicts abnormalities and then generates targeted descriptions for each. Evaluation on a public dataset demonstrates significant improvements in report quality and clinical relevance. We extend our work by conducting an ablation study to demonstrate its effectiveness.}

\section{Introduction}
\label{sec:introduction}
Three-dimensional Computed Tomography (3D CT) scans have emerged as indispensable tools in medical imaging~\cite{singh_3d_2020}. Given the rapidly increasing number of scans to analyze~\cite{hess_trends_2014, broder_increasing_2006, bellolio_increased_2017}, the time-consuming nature of the task~\cite{goergen_evidence-based_2013} and the important demand for specialized radiological expertise in numerous healthcare systems~\cite{bastawrous_improving_2017, rimmer_radiologist_2017}, computed-aided diagnosis and automating report generation have emerged as an active research area ~\cite{ranschaert_optimization_2021, liu_benchmarking_2024, hamamci_ct2rep_2024}. Whether dealing with 2D~\cite{johnson_mimic-cxr_2019, irvin_chexpert_2019} or 3D~\cite{draelos_machine-learning-based_2021, hamamci_foundation_2024} images, numerous deep learning methods have been introduced to support healthcare professionals in tasks such as anomaly classification~\cite{draelos_machine-learning-based_2021, noauthor_deep-chest_nodate}, detection~\cite{rehman_review_2023, dubey_computer-aided_2020, kim_3d_2024}, segmentation~\cite{navab_deep_2015, mai_systematic_2023, ilesanmi_reviewing_2024}, and report generation~\cite{liu_benchmarking_2024, hamamci_ct2rep_2024, tanida_interactive_2023, chen_generating_2022}. Indeed, the availability of public datasets for 2D radiology images~\cite{irvin_chexpert_2019, johnson_mimic-cxr_2019} has facilitated the development of such methods~\cite{tanida_interactive_2023, chen_generating_2022, jing_show_2019, jing_automatic_2018, rio-torto_parameter-efficient_nodate, stock_generalist_2024}. Recently, CT2Rep~\cite{hamamci_ct2rep_2024} has been presented as the first 3D method for automated report generation, trained and evaluated on the public CT-RATE dataset~\cite{hamamci_foundation_2024} of CT examination. CT2Rep, uses an Encoder-Decoder model to generate reports directly from 3D CT scans. However, this end-to-end approach may lead to incomplete reports~\cite{bastawrous_improving_2017}. Reading a CT study and composing a report~\cite{ridley_guide_2002}, a radiologist detects abnormalities and then provides descriptions for each identified abnormality. Inspired by the radiologists' workflow, we propose a method that first predicts anomalies and then generates a specific sentence for each anomaly. Our contributions are as follows:

\begin{itemize}[noitemsep,topsep=0pt,parsep=0pt,partopsep=0pt]
    \item We introduce a new abnormality-based sentence generation model that enhances chest 3D CT scan report generation performance, achievable with limited computational resources (single GPU, 24-hour training time).
    \item We leverage a pre-trained language model with biomedical domain knowledge that we condition to generate a sentence for each abnormality.
    \item We evaluate the model on a public dataset and we enhance our findings with an ablation study to show the effectiveness of each module.
\end{itemize} 

\section{Related work}
\label{sec:related_work}

\subsection{2D Radiology report generation}

Image captioning has emerged as a crucial task across diverse research domains~\cite{gurari_vizwiz_2018, xue_multi-task_2024, vinyals_show_2015, zhang_remote_nodate, chen_unit3d_2022}. This field has garnered significant attention in recent years in recent years~\cite{wang_overview_2020, lopez-sanchez_supervised_2023, gaurav_survey_2021}, with applications in automated medical image reporting~\cite{pang_survey_2023}. Early supervised methods for 2D radiology report generation~\cite{johnson_mimic-cxr_2019, irvin_chexpert_2019} primarily used Encoder-Decoder architectures~\cite{jing_show_2019, jing_automatic_2018, noauthor_ybrid_nodate} . The features of the input image are extracted using an encoder, either a CNN~\cite{oshea_introduction_2015} or a Vision Transformer~\cite{dosovitskiy_image_2021, vaswani_attention_2023}. These image-level features are then passed to a decoder, corresponding to a RNN~\cite{sherstinsky_fundamentals_2020} or Transformer Decoder~\cite{chen_generating_2022, rio-torto_parameter-efficient_nodate}, which predicts the corresponding report. More recent methods introduce relational memory~\cite{chen_generating_2022}, pathology tags~\cite{you_aligntransformer_2022}, or reinforcement learning~\cite{noauthor_large_nodate} to improve the quality of generated reports. However, these methods generate the entire report only from the radiology image. Given the complexity and the inherent variability in radiological findings, these methods tend to produce incomplete reports~\cite{bastawrous_improving_2017}. Guided methods were then introduced to facilitate report generation~\cite{nooralahzadeh_progressive_2021, liu_clinically_2019}. 
Descriptive sentences specific to identified regions or predicted abnormalities can be generated through the integration of an anatomical region detection or abnormality classification module~\cite{tanida_interactive_2023}.

\begin{figure*}[t]
        \centering
        \includegraphics[width=0.95\linewidth]{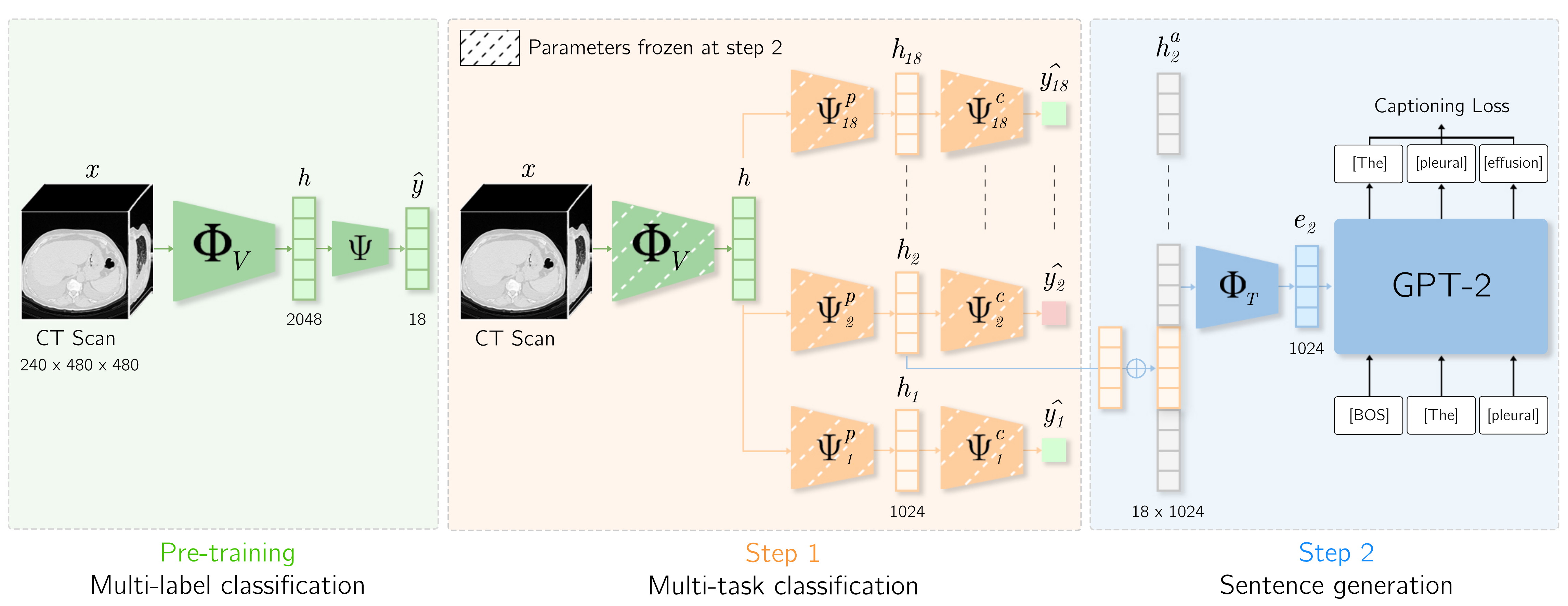}
        %\vspace{-1em}
        \vspace{4pt}
        \caption{Overview of the method. \textbf{Pre-training.} The input volume $x$ is passed through a visual extractor $\Phi_{V}$ (either CT-ViT~\cite{hamamci_generatect_2024} or  CT-Net~\cite{draelos_machine-learning-based_2021}) to extract an embedding $h$. $h$ is then given to classification head $\Psi$ which predicts the logit vector $\hat{y}$. \textbf{Step 1.} $h$ is fed into 18 projection heads ($\Psi^{p}_{i}, i \in \{1, \ldots, 18\}$) followed by small classification heads ($\Psi^{c}_{i}, i \in \{1, \ldots, 18\}$), one for each label. This second step enables to obtain an embedding $h_{i}$ (and then $h^{a}_{i}$) specific to each label. \textbf{Step 2.} If a label indexed by $i$ is predicted as abnormal by its corresponding classification head $\Psi^{c}_{i}$, the associated embedding $h^{a}_{i}$ is transformed by a lightweight MLP $\Phi_{T}(e_{i})$ to obtain $e_{i}$. An abnormality-specific description is generated from a pre-trained GPT-2 using $e_{i}$.}
        \label{fig:teaser}
\end{figure*}

\subsection{3D Radiology report generation}
New challenges arise with 3D imaging~\cite{bai_m3d_2024, wu_towards_2023}. Compared to 2D images, this modality of data introduces new difficulties due to the limited availability of public datasets~\cite{hamamci_ct2rep_2024} in this field and the significant computational resources needed~\cite{bai_m3d_2024, wu_towards_2023}.
Abnormality detection ~\cite{kim_3d_2024, djahnine_detection_2024} and classification~\cite{draelos_machine-learning-based_2021, hamamci_foundation_2024} for 3D CT scans have been the subject of research efforts in recent years. The Transformer-based visual feature extractor CT-ViT~\cite{hamamci_generatect_2024} has been introduced to deal with 3D CT scan efficiently, enabling the development of CT2Rep~\cite{hamamci_ct2rep_2024}. CT2Rep is the first method for automatic 3D CT scan radiology report generation, trained and evaluated on the CT-RATE public dataset~\cite{hamamci_foundation_2024}. CT2Rep is an auto-regressive model based on an encoder-decoder architecture~\cite{chen_generating_2022}, where visual features are extracted from the 3D CT scan using CT-ViT and then given to a Transformer Decoder~\cite{chen_generating_2022, vaswani_attention_2023} that generates the report. 
To improve the completeness of generated reports, our approach builds upon previous work~\cite{tanida_interactive_2023, wang_inclusive_2022} and introduces a step for abnormality classification to guide report generation based on previously predicted anomalies. This method offers greater control over the generation process.

\section{Dataset}
\label{sec:dataset}

We use the CT-RATE public dataset~\cite{hamamci_foundation_2024} to train and evaluate our method. CT-RATE contains reconstructed non-contrast 3D chest CT volumes, corresponding reports and 18 distinct types of abnormalities extracted from reports using RadBERT~\cite{hamamci_foundation_2024, yan_radbert_2022}. Train set has 17,799 unique patients (34,781 volumes), validation set has 1,314 patients (3,075 volumes) and test set has 1,314 patients (3,039 volumes). Following~\cite{draelos_machine-learning-based_2021, hamamci_foundation_2024}, all volumes are either center-cropped or padded to achieve a resolution of $240 \times 480 \times 480$ with a in-slice spacing of $0.75$ mm and $1.5$ mm in the z-axis. Hounsfield Unit (HU)~\cite{denotter_hounsfield_2024} values are clipped between $-1000$ to $+200$ before normalization to [-1, 1].

\section{Method}
\label{sec:method}

As illustrated in Figure~\ref{fig:teaser}, we employ a two-stage approach for anomaly detection and description generation. Initially, we use a visual feature extractor pre-trained on a multi-label classification task. In the first stage, we perform multi-task learning with one classification head per anomaly. If an anomaly is detected, its associated vector representation is then passed to the second stage. Here, a pre-trained GPT-2 model generates a descriptive text of the identified anomaly.

\subsection{Multi-label classification pre-training}
\label{section:step1}
The model receives a volume $x \in \mathbb{R}^{240 \times 480 \times 480}$ as input. This volume is passed to a visual feature extractor $\Phi_{V}$, either CT-Net~\cite{draelos_machine-learning-based_2021} or CT-ViT~\cite{hamamci_generatect_2024}. CT-Net consists of ResNet~\cite{he_deep_2015} modules pre-trained on ImageNet~\cite{noauthor_imagenet_nodate} followed by a 3D convolutional network~\cite{tran_learning_2015}. CT-ViT~\cite{hamamci_generatect_2024} is a Vision Transformer~\cite{dosovitskiy_image_2021} based on the attention mechanism~\cite{vaswani_attention_2023} computed from 3D patches extracted from the initial volume. From the initial volume $x$, both CT-Net and CT-ViT yield a vector representation $h \in \mathbb{R}^{2048}$, such that:
\begin{equation}
\label{eq:embedding_h}
h = \Phi_{V}(x).
\end{equation}

Then, $h$ is fed into a classification head $\Psi$ to predict $\hat{y} \in \mathbb{R}^{18}$, as follows:
\begin{equation}
\label{eq:prediction_step1}
\hat{y} = \Psi(h)\, ,
\end{equation}

where the components of the logit vector $\mathbf{\hat{y}}$ represent the anomaly scores for their respective class labels.
The model is trained using a binary cross-entropy loss function~\cite{bogatinovski_comprehensive_2022, good_rational_1952}. This initial step facilitates the training of a visual encoder $\Phi_{V}$ to extract salient visual features that enable accurate anomaly prediction.

\subsection{Step 1: Multi-task single label classification}
\label{section:step2}
The multi-label classification head $\Psi$ of the pre-trained network is replaced by 18 independent projection heads~\cite{chen_simple_2020}, denoted as $\Psi^{p}_{i}$ ($i \in \{1, \ldots, 18\}$), followed by classification heads denoted as $\Psi^{c}_{i}$ ($i \in \{1, \ldots, 18\}$). The feature vector $h$ is first passed through each projection head $\Psi^{p}_{i}$ to obtain a unique embedding noted $h_{i} \in \mathbb{R}^{1024}$ per class label, such that:
\begin{equation}
\label{eq:embedding_e}
h_{i} = \Psi^{p}_{i}(h) \ \forall \ \ i \in \{1, \ldots, 18\}\, .
\end{equation}

Subsequently, each embedding $h_{i}$ is fed into its corresponding classification head $\Psi^{c}_{i}$. The classification head outputs an anomaly score $\hat{y}_{i} \in \mathbb{R}$ and employs a binary cross-entropy loss $\mathcal{L}_{i} \in \mathbb{R}$ as the objective function. The prediction step can be formalized as:
\begin{equation}
\label{eq:prediction_step2}
\hat{y}_{i} = \Psi^{c}_{i}(h_{i}) \ \forall \ i \in \{1, \ldots, 18\}.
\end{equation}
Here, $\hat{y}_{i}$ represents the predicted anomaly score for class $i$. The binary cross-entropy loss is then computed for each class, such that:
\begin{equation}
\label{eq:loss}
\mathcal{L}_{i}(\hat{y}_{i}, y_{i}) = -[y_{i} \log(\hat{y}_{i}) + (1-y_{i}) \log(1-\hat{y}_{i})],
\end{equation}
where $y_{i} \in \{0,1\}$ denotes the ground-truth label for class $i$.

For each projection $\Psi^{p}_{i}$ and classification $\Psi^{c}_{i}$ head, back-propagation~\cite{lecun_theoretical_2001} of the corresponding parameters is performed solely based on the associated loss $\mathcal{L}_{i}$. For the rest of the neural network, back-propagation is carried out using the sum of these losses, denoted as $\mathcal{L} \in \mathbb{R}$. Introducing this supervised multi-task learning step enhances classification performance~\cite{crawshaw_multi-task_2020} and enables the extraction of a unique vector representation $h_{i}$ for each abnormality $i$, conditioning sentence generation per abnormality in the subsequent step.

\subsection{Step 2: Abnormality-based sentence generation}
\label{section:step3}

\textbf{Abnormality selection} A description is generated for each abnormality predicted by the classification module. During the training of this generation module, we consider only the labels correctly predicted as abnormal by the classification heads. During inference, we only consider labels predicted as abnormal by the corresponding classification heads. A label $i$ is classified as abnormal if its predicted score $\hat{y_{i}}$ exceeds a threshold computed from the validation set, which maximizes the F1-Score~\cite{powers_evaluation_2020, sasaki_truth_nodate, litjens_survey_2017}.

\noindent \textbf{Abnormality feature extraction} For each label indexed by $i$ and predicted as abnormal by its corresponding classification head $\Psi^{c}_{i}$, we extract and expend the corresponding embedding $h_{i} \in \mathbb{R}^{1024}$ to obtain a vector $h^{a}_{i} \in \mathbb{R}^{18\times1024}$. This vector $h^{a}_{i}$ contains zeros for all components associated with other abnormalities, while retaining the original embedding $h_{i}$. Using a lightweight MLP noted $\Phi_{T}$, we project $h^{a}_{i}$ from the visual latent space to the textual latent space, yielding $e_{i} \in \mathbb{R}^{1024}$, such that:
\begin{equation}
\label{eq:embedding_eT}
e_{i} = \Phi_{T}(h^{a}_{i})\, ,
\end{equation}
\begin{equation*}
 \begin{aligned}
\label{eq:embedding_eT_zeros}
\text{where } h^{a}_{i} = [\vec{0}, \ \ldots, \ \vec{0}, \ h_{i}, \ \vec{0}, \ \ldots, \ \vec{0}] \, ,\\
\text{and } \vec{0} \text{ denotes a zero vector of dimension 1024.}
\end{aligned}
\end{equation*}
%\begin{equation}
%\label{eq:zeros}
%\text{and } \vec{0} \text{ denotes a zero vector of dimension 1024.}
%\end{equation}

\noindent This allows for a focused representation of the predicted abnormality within the multi-abnormality embedding space, while maintaining the context of the original embedding.

\noindent \textbf{Sentence generation} For each label $i$ predicted as abnormal, we generate a descriptive sentence based on its associated embedding $e_{i}$. Inspired by prior work~\cite{tanida_interactive_2023}, the GPT-2 Medium Language model~\cite{radford_language_nodate} fine-tuned on PubMed abstracts~\cite{papanikolaou_dare_2020} is used for the Decoder part. GPT-2 is an auto-regressive model grounded in attention mechanism~\cite{vaswani_attention_2023} where each embedded token only considers the contextual information provided by preceding tokens. Instead of using the conventional attention mechanism, the language model is conditioned on anomaly prediction by employing pseudo self-attention~\cite{ziegler_encoder-agnostic_2019}, noted $PS$. This approach injects the features associated with abnormality classification into the self-attention mechanism of GPT-2, such that:
\begin{equation}
\label{eq:embedding}
PS(e_{i}, Y) = \text{softmax}\left( (Y W_{q}) \begin{bmatrix}
e_{i} w_{k} \\
Y W_{k}
\end{bmatrix}^{T} \right) \begin{bmatrix}
e_{i} w_{v} \\
Y W_{v}
\end{bmatrix},
\end{equation}
\noindent where $w_{k}$ and $w_{v}$ are the projection parameters for key and value vectors~\cite{vaswani_attention_2023} for $e_{i}$; $W_k$, $W_q$ and $W_v$ are the projection parameters for key, query and value for $Y$ which are textual embedded tokens. During training, the model is trained on a next token prediction task~\cite{bengio_neural_nodate} using a cross-entropy loss function~\cite{hamamci_ct2rep_2024, tanida_interactive_2023, chen_generating_2022}. All parameters are frozen except for those of GPT-2. During inference, only the CT scan and the [BOS] token~\cite{radford_language_nodate} are fed into the model, which generates a description for each predicted abnormality. 

\noindent \textbf{Report generation} The final report is obtained by concatenating the generated sentences associated with each predicted anomaly.

%%% WITH STD
\begin{table*}[h]
\centering
\begin{tabular}{l c c c c c c c}
\toprule%
& \multicolumn{4}{@{}c@{}}{\small \textbf{NLG Metrics}} & \multicolumn{3}{@{}c@{}}{\small \textbf{CE Metrics}} \\
%\cmidrule{2-6}\cmidrule{8-10}
\cmidrule(lr){2-5} \cmidrule(lr){6-8}
\textbf{\small Method} & \small \textbf{METEOR} & \small \textbf{ROUGE\textsubscript{L}} & \small \textbf{BERT} & \small \textbf{BART} & \small \textbf{P} & \small \textbf{R} & \small \textbf{F1}\\
\toprule
\small CT2Rep~\cite{hamamci_ct2rep_2024} & \small $0.173 \text{\scriptsize $\pm 0.002$}$ & \small $0.243 \text{\scriptsize $\pm 0.002$}$ & \small $0.865 \text{\scriptsize $\pm 0.001$}$ & \small $-3.314 \text{\scriptsize $\pm 0.085$}$ & \small $41.63 \text{\scriptsize $\pm 2.01$}$ & \small $38.12 \text{\scriptsize $\pm 0.88$}$ & \small $36.73 \text{\scriptsize $\pm 0.76$}$\\
\hline
\rowcolor[gray]{0.93} 
\small CT-AGRG \makebox[0.05cm][l]{} w/ CT-ViT & \small $0.190 \text{\scriptsize $\pm 0.004$}$ & \small $0.270 \text{\scriptsize $\pm 0.002$}$ & \small $0.864 \text{\scriptsize $\pm 0.001$}$ & \small $-2.936 \text{\scriptsize $\pm 0.040$}$ & \small $37.75 \text{\scriptsize $\pm 0.46$}$ & \small $55.35 \text{\scriptsize $\pm 0.91$}$ & \small $42.13 \text{\scriptsize $\pm 0.23$}$\\
\rowcolor[gray]{0.93} 
\small CT-AGRG \makebox[0.05cm][l]{} w/ CT-Net & \small $\mathbf{0.196} \text{\scriptsize $\pm 0.008$}$ & \small $\mathbf{0.280}\text{\scriptsize $\pm 0.003$}$ & \small $\mathbf{0.867} \text{\scriptsize $\pm 0.001$}$ & \small $\mathbf{-2.714} \text{\scriptsize $\pm 0.179$}$ & \small $\mathbf{45.74} \text{\scriptsize $\pm 0.35$}$ & \small $\mathbf{62.59} \text{\scriptsize $\pm 0.14$}$ & \small $\mathbf{50.11} \text{\scriptsize $\pm 0.15$}$\\
\toprule
\end{tabular}

\centering
\begin{tabular}{c c c c | c c c c}
\multicolumn{1}{c}{\small \textbf{\textit{Ablation study}}} \\
\hline
\begin{tabular}{@{}c@{}}\small Multi-label \\ \small classification\end{tabular} & 
\begin{tabular}{@{}c@{}}\small Multi-task \\ \small classification\end{tabular} & 
\begin{tabular}{@{}c@{}}\small Multi-Abnormality \\ \small Embedding\end{tabular} & 
\begin{tabular}{@{}c@{}}\small Language \\ \small model\end{tabular} & 
\small \textbf{BLEU-4} & \small \textbf{METEOR} & \small \textbf{R} & \small \textbf{F1}\\
\hline

\small \scalecheck &  &  & \small \scalecheck & \small $0.083 \text{\scriptsize $\pm 0.003$}$ & \small $0.127 \text{\scriptsize $\pm 0.008$}$ & \small $19.05 \text{\scriptsize $\pm 0.19$}$ & \small $23.80 \text{\scriptsize $\pm 0.10$}$\\
\small \scalecheck & \small \scalecheck  & & \small \scalecheck & \small $0.154 \text{\scriptsize $\pm 0.008$}$ & \small $0.186 \text{\scriptsize $\pm 0.005$}$ & \small $56.56 \text{\scriptsize $\pm 0.12$}$ & \small $46.11 \text{\scriptsize $\pm 0.13$}$\\
\small \scalecheck  &  & \small \scalecheck & \small \scalecheck & \small $0.146 \text{\scriptsize $\pm 0.005$}$ & \small $0.189 \text{\scriptsize $\pm 0.007$}$ & \small $58.89 \text{\scriptsize $\pm 0.12$}$ & \small $47.22 \text{\scriptsize $\pm 0.11$}$\\

\rowcolor[gray]{0.93} 
\small \scalecheck & \small \scalecheck & \small \scalecheck & \small \scalecheck & \small $\mathbf{0.172}\text{\scriptsize $\pm 0.007$}$ & \small $\mathbf{0.196}\text{\scriptsize $\pm 0.008$}$ & \small $\mathbf{62.59}\text{\scriptsize $\pm 0.14$}$ & \small $\mathbf{50.11}\text{\scriptsize $\pm 0.15$}$\\

\toprule
\end{tabular}
\vspace{1em}
\caption{\textbf{Quantitative evaluation on the test set} with Natural Language Generation (NLG) metrics and Clinical Efficacy (CE) metrics. Reported mean and standard deviation metrics were computed over five independent runs. Our model outperforms the previous method both on NLG and CE metrics. \textbf{\textit{Ablation study:}} The ablation study on CT-AGRG was conducted with CT-Net~\cite{draelos_machine-learning-based_2021} as visual feature extractor, since it yields the best results. It highlights the importance of each module.}\label{tab:quantitative}
\end{table*}

\section{Experimental setup}
\label{sec:experimental_setup}

\textbf{Training details.} For the first and second steps, the model was trained with a batch size of 4, using the Adam~\cite{kingma_adam_2017} optimizer for 40 and 2 epochs, with a learning rate of $10^{-4}$ and $10^{-5}$, respectively. For the third step, training was conducted for 50 epochs with a batch size of 64, using the AdamW~\cite{loshchilov_decoupled_2019} optimizer with a learning rate of $5 \times 10^{-5}$. The training required a GPU with 48GB of memory.

\noindent \textbf{Generation mode.} Generated sentences were limited to 60 tokens~\cite{levy_same_2024}, typically the maximum observed per sentence in the report corpus. At inference, we use the Beam Search~\cite{lowerre_harpy_nodate, freitag_beam_2017, meister_best-first_2022} algorithm as generation mode~\cite{jurafsky2009speech, li_pretrained_2022, vijayakumar_diverse_2018, noauthor_hierarchical_nodate}. Beam size is set to 4, aligning with prior work and balancing output quality with computational efficiency.~\cite{hamamci_ct2rep_2024, tanida_interactive_2023, chen_generating_2022}.
 
\section{Results}
\label{sec:results}

\subsection{Quantitative results}
Abnormalities mentioned in the generated report are extracted with the RadBERT labeler~\cite{hamamci_foundation_2024, yan_radbert_2022}. These predicted anomalies are compared against ground truth labels to compute Clinical Efficacy (CE) metrics: Precision, Recall, and F1-Score~\cite{van1979information}. We also evaluate our captioning model using widely used~\cite{hamamci_ct2rep_2024, tanida_interactive_2023, chen_generating_2022, rio-torto_parameter-efficient_nodate} Natural Language Generation (NLG) metrics: BLEU-4~\cite{papineni_bleu_2002} for n-gram precision; METEOR~\cite{lavie_meteor_2009} for alignment with stemming and synonymy; ROUGE~\cite{noauthor_text_2004} for overlap focusing on recall; BERT-Score~\cite{zhang_bertscore_2020} for token-level contextual embedding similarity; and BART-Score~\cite{yuan_bartscore_2021} for sequence-level likelihood assessment. This combination captures lexical, semantic, and generative aspects of caption quality, providing a comprehensive evaluation of our model's performance~\cite{kayser_e-vil_2021, kayser_explaining_2022}. Table~\ref{tab:quantitative} demonstrates that CT-AGRG achieves significantly better results than prior work. With CT-ViT and CT-Net as visual feature extractors, we achieve an average Recall of 55.35 and 62.59 ($\Delta$+45.20\% and $\Delta$+64.19\% increase compared to CT2Rep~\cite{hamamci_ct2rep_2024}), and an average F1-Score of 42.13 and 50.11, respectively ($\Delta$+14.70\% and $\Delta$+36.43\% increase). CT-AGRG achieves a BART-Score~\cite{yuan_bartscore_2021} of -2.714, outperforming CT2Rep~\cite{hamamci_ct2rep_2024} (-3.314, $\Delta$+18.11\% increase), indicating enhanced abnormality detection capabilities and improved semantic relevance in generated captions. 

\subsection{Ablation study}

As shown in Table~\ref{tab:quantitative}, we conduct an ablation study to evaluate each component of our CT-AGRG model. Introducing multi-task classification significantly improves F1-Score from 23.80 to 46.11 ($\Delta$+93.76\% increase). Augmenting the latent space further enhances performance, increasing Recall from 56.56 to 58.89 ($\Delta$+4.12\% increase) and F1-Score from 46.11 to 47.22 ($\Delta$+2.41\% increase). NLG metrics show notable improvements, with BLEU-4~\cite{papineni_bleu_2002} and METEOR~\cite{lavie_meteor_2009} scores increasing by 0.089 and 0.069, respectively, indicating enhanced sentence variability while maintaining semantic fidelity to ground truth reports.

\section{Conclusion}
\label{sec:conclusion}

In this paper, we have presented an effective method for guided report generation. Our approach, based on conditional description generation for each abnormality, outperforms state-of-the-art methods while offering a high level of control over the generated report. Future work could focus on adapting the approach to require less supervision, leveraging other modalities or region-specific data to further enhance the report generation process. Access CT-AGRG implementation at \url{https://github.com/theodpzz/ct-agrg}.

\newpage

\section{Acknowledgments}
\label{sec:acknowlegments}
We acknowledge Hamamci et al. 2024~\cite{hamamci_foundation_2024} for making the CT-RATE dataset available.

\bibliographystyle{IEEEbib}
\bibliography{references}

@book{van1979information,
  title={Information Retrieval},
  author={van Rijsbergen, C. J.},
  year={1979},
  publisher={Butterworth-Heinemann}
}

@book{jurafsky2009speech,
  title={Speech and Language Processing: An Introduction to Natural Language Processing, Computational Linguistics, and Speech Recognition},
  author={Jurafsky, Daniel and Martin, James H},
  year={2009},
  edition={2},
  publisher={Prentice Hall},
  address={Upper Saddle River, NJ},
  isbn={978-0131873216}
}

@article{litjens_survey_2017,
	title = {A survey on deep learning in medical image analysis},
	volume = {42},
	issn = {1361-8415},
	url = {https://www.sciencedirect.com/science/article/pii/S1361841517301135},
	doi = {https://doi.org/10.1016/j.media.2017.07.005},
	journal = {Medical Image Analysis},
	author = {Litjens, Geert and Kooi, Thijs and Bejnordi, Babak Ehteshami and Setio, Arnaud Arindra Adiyoso and Ciompi, Francesco and Ghafoorian, Mohsen and Laak, Jeroen A. W. M. van der and Ginneken, Bram van and Sánchez, Clara I.},
	year = {2017},
	pages = {60--88},
}

@incollection{navab_deep_2015,
	address = {Cham},
	title = {Deep {Learning} and {Structured} {Prediction} for the {Segmentation} of {Mass} in {Mammograms}},
	volume = {9349},
	isbn = {978-3-319-24552-2 978-3-319-24553-9},
	url = {http://link.springer.com/10.1007/978-3-319-24553-9_74},
	language = {en},
	urldate = {2024-06-17},
	booktitle = {Medical {Image} {Computing} and {Computer}-{Assisted} {Intervention} -- {MICCAI} 2015},
	publisher = {Springer International Publishing},
	author = {Dhungel, Neeraj and Carneiro, Gustavo and Bradley, Andrew P.},
	editor = {Navab, Nassir and Hornegger, Joachim and Wells, William M. and Frangi, Alejandro},
	year = {2015},
	doi = {10.1007/978-3-319-24553-9_74},
	note = {Series Title: Lecture Notes in Computer Science},
	pages = {605--612},
}

@misc{irvin_chexpert_2019,
	title = {{CheXpert}: {A} {Large} {Chest} {Radiograph} {Dataset} with {Uncertainty} {Labels} and {Expert} {Comparison}},
	shorttitle = {{CheXpert}},
	url = {http://arxiv.org/abs/1901.07031},
	doi = {10.48550/arXiv.1901.07031},
	urldate = {2024-06-17},
	publisher = {arXiv},
	author = {Irvin, Jeremy and Rajpurkar, Pranav and Ko, Michael and Yu, Yifan and Ciurea-Ilcus, Silviana and Chute, Chris and Marklund, Henrik and Haghgoo, Behzad and Ball, Robyn and Shpanskaya, Katie and Seekins, Jayne and Mong, David A. and Halabi, Safwan S. and Sandberg, Jesse K. and Jones, Ricky and Larson, David B. and Langlotz, Curtis P. and Patel, Bhavik N. and Lungren, Matthew P. and Ng, Andrew Y.},
	month = jan,
	year = {2019},
	note = {arXiv:1901.07031 [cs, eess]},
}

@misc{hamamci_ct2rep_2024,
	title = {{CT2Rep}: {Automated} {Radiology} {Report} {Generation} for {3D} {Medical} {Imaging}},
	shorttitle = {{CT2Rep}},
	url = {http://arxiv.org/abs/2403.06801},
	doi = {10.48550/arXiv.2403.06801},
	urldate = {2024-06-17},
	publisher = {arXiv},
	author = {Hamamci, Ibrahim Ethem and Er, Sezgin and Menze, Bjoern},
	month = mar,
	year = {2024},
	note = {arXiv:2403.06801 [cs, eess]},
}

@misc{hamamci_generatect_2024,
	title = {{GenerateCT}: {Text}-{Conditional} {Generation} of {3D} {Chest} {CT} {Volumes}},
	shorttitle = {{GenerateCT}},
	url = {http://arxiv.org/abs/2305.16037},
	doi = {10.48550/arXiv.2305.16037},
	urldate = {2024-06-17},
	publisher = {arXiv},
	author = {Hamamci, Ibrahim Ethem and Er, Sezgin and Sekuboyina, Anjany and Simsar, Enis and Tezcan, Alperen and Simsek, Ayse Gulnihan and Esirgun, Sevval Nil and Almas, Furkan and Dogan, Irem and Dasdelen, Muhammed Furkan and Prabhakar, Chinmay and Reynaud, Hadrien and Pati, Sarthak and Bluethgen, Christian and Ozdemir, Mehmet Kemal and Menze, Bjoern},
	month = mar,
	year = {2024},
	note = {arXiv:2305.16037 [cs]},
}

@misc{hamamci_foundation_2024,
	title = {A foundation model utilizing chest {CT} volumes and radiology reports for supervised-level zero-shot detection of abnormalities},
	url = {http://arxiv.org/abs/2403.17834},
	doi = {10.48550/arXiv.2403.17834},
	urldate = {2024-06-17},
	publisher = {arXiv},
	author = {Hamamci, Ibrahim Ethem and Er, Sezgin and Almas, Furkan and Simsek, Ayse Gulnihan and Esirgun, Sevval Nil and Dogan, Irem and Dasdelen, Muhammed Furkan and Wittmann, Bastian and Simsar, Enis and Simsar, Mehmet and Erdemir, Emine Bensu and Alanbay, Abdullah and Sekuboyina, Anjany and Lafci, Berkan and Ozdemir, Mehmet K. and Menze, Bjoern},
	month = mar,
	year = {2024},
	note = {arXiv:2403.17834 [cs]},
}

@inproceedings{tanida_interactive_2023,
	title = {Interactive and {Explainable} {Region}-guided {Radiology} {Report} {Generation}},
	url = {http://arxiv.org/abs/2304.08295},
	doi = {10.1109/CVPR52729.2023.00718},
	urldate = {2024-06-17},
	booktitle = {2023 {IEEE}/{CVF} {Conference} on {Computer} {Vision} and {Pattern} {Recognition} ({CVPR})},
	author = {Tanida, Tim and Müller, Philip and Kaissis, Georgios and Rueckert, Daniel},
	month = jun,
	year = {2023},
	note = {arXiv:2304.08295 [cs]},
	pages = {7433--7442},
}

@article{draelos_machine-learning-based_2021,
	title = {Machine-learning-based multiple abnormality prediction with large-scale chest computed tomography volumes},
	volume = {67},
	issn = {1361-8415},
	url = {https://www.sciencedirect.com/science/article/pii/S1361841520302218},
	doi = {10.1016/j.media.2020.101857},
	urldate = {2024-06-17},
	journal = {Medical Image Analysis},
	author = {Draelos, Rachel Lea and Dov, David and Mazurowski, Maciej A. and Lo, Joseph Y. and Henao, Ricardo and Rubin, Geoffrey D. and Carin, Lawrence},
	month = jan,
	year = {2021},
	pages = {101857},
}

@article{djahnine_detection_2024,
	title = {Detection and severity quantification of pulmonary embolism with {3D} {CT} data using an automated deep learning-based artificial solution},
	volume = {105},
	issn = {2211-5684},
	url = {https://www.sciencedirect.com/science/article/pii/S2211568423001808},
	doi = {10.1016/j.diii.2023.09.006},
	number = {3},
	urldate = {2024-06-17},
	journal = {Diagnostic and Interventional Imaging},
	author = {Djahnine, Aissam and Lazarus, Carole and Lederlin, Mathieu and Mulé, Sébastien and Wiemker, Rafael and Si-Mohamed, Salim and Jupin-Delevaux, Emilien and Nempont, Olivier and Skandarani, Youssef and De Craene, Mathieu and Goubalan, Segbedji and Raynaud, Caroline and Belkouchi, Younes and Afia, Amira Ben and Fabre, Clement and Ferretti, Gilbert and De Margerie, Constance and Berge, Pierre and Liberge, Renan and Elbaz, Nicolas and Blain, Maxime and Brillet, Pierre-Yves and Chassagnon, Guillaume and Cadour, Farah and Caramella, Caroline and Hajjam, Mostafa El and Boussouar, Samia and Hadchiti, Joya and Fablet, Xavier and Khalil, Antoine and Talbot, Hugues and Luciani, Alain and Lassau, Nathalie and Boussel, Loic},
	month = mar,
	year = {2024},
	pages = {97--103},
}

@article{rehman_review_2023,
	title = {Review on chest pathogies detection systems using deep learning techniques},
	volume = {56},
	issn = {1573-7462},
	url = {https://doi.org/10.1007/s10462-023-10457-9},
	doi = {10.1007/s10462-023-10457-9},
	language = {en},
	number = {11},
	urldate = {2024-06-17},
	journal = {Artificial Intelligence Review},
	author = {Rehman, Arshia and Khan, Ahmad and Fatima, Gohar and Naz, Saeeda and Razzak, Imran},
	month = nov,
	year = {2023},
	pages = {12607--12653},
}

@article{goergen_evidence-based_2013,
	title = {Evidence-based guideline for the written radiology report: methods, recommendations and implementation challenges},
	volume = {57},
	issn = {1754-9485},
	shorttitle = {Evidence-based guideline for the written radiology report},
	doi = {10.1111/1754-9485.12014},
	language = {eng},
	number = {1},
	journal = {Journal of Medical Imaging and Radiation Oncology},
	author = {Goergen, Stacy K. and Pool, Felicity J. and Turner, Tari J. and Grimm, Jane E. and Appleyard, Mark N. and Crock, Carmel and Fahey, Michael C. and Fay, Michael F. and Ferris, Nicholas J. and Liew, Susan M. and Perry, Richard D. and Revell, Ann and Russell, Grant M. and Wang, Shih-Chang S. C. and Wriedt, Christian},
	month = feb,
	year = {2013},
	pmid = {23374546},
	pages = {1--7},
}

@article{bastawrous_improving_2017,
	title = {Improving {Patient} {Safety}: {Avoiding} {Unread} {Imaging} {Exams} in the {National} {VA} {Enterprise} {Electronic} {Health} {Record}},
	volume = {30},
	issn = {1618-727X},
	shorttitle = {Improving {Patient} {Safety}},
	doi = {10.1007/s10278-016-9937-2},
	language = {eng},
	number = {3},
	journal = {Journal of Digital Imaging},
	author = {Bastawrous, Sarah and Carney, Benjamin},
	month = jun,
	year = {2017},
	pmid = {28050717},
	pmcid = {PMC5422229},
	pages = {309--313},
}

@article{rimmer_radiologist_2017,
	title = {Radiologist shortage leaves patient care at risk, warns royal college},
	volume = {359},
	issn = {1756-1833},
	doi = {10.1136/bmj.j4683},
	language = {eng},
	journal = {BMJ (Clinical research ed.)},
	author = {Rimmer, Abi},
	month = oct,
	year = {2017},
	pmid = {29021184},
	pages = {j4683},
}

@misc{johnson_mimic-cxr_2019,
	title = {The {MIMIC}-{CXR} {Database}},
	url = {https://physionet.org/content/mimic-cxr/},
	doi = {10.13026/C2JT1Q},
	urldate = {2024-06-17},
	publisher = {physionet.org},
	author = {Johnson, Alistair E. W. and Pollard, Tom and Mark, Roger and Berkowitz, Seth and Horng, Steven},
	year = {2019},
}

@article{kim_3d_2024,
	title = {{3D} unsupervised anomaly detection through virtual multi-view projection and reconstruction: {Clinical} validation on low-dose chest computed tomography},
	volume = {236},
	issn = {0957-4174},
	shorttitle = {{3D} unsupervised anomaly detection through virtual multi-view projection and reconstruction},
	url = {https://www.sciencedirect.com/science/article/pii/S0957417423016676},
	doi = {10.1016/j.eswa.2023.121165},
	urldate = {2024-06-17},
	journal = {Expert Systems with Applications},
	author = {Kim, Kyungsu and Oh, Seong Je and Lee, Ju Hwan and Chung, Myung Jin},
	month = feb,
	year = {2024},
	pages = {121165},
}

@misc{dosovitskiy_image_2021,
	title = {An {Image} is {Worth} 16x16 {Words}: {Transformers} for {Image} {Recognition} at {Scale}},
	shorttitle = {An {Image} is {Worth} 16x16 {Words}},
	url = {http://arxiv.org/abs/2010.11929},
	doi = {10.48550/arXiv.2010.11929},
	urldate = {2024-06-17},
	publisher = {arXiv},
	author = {Dosovitskiy, Alexey and Beyer, Lucas and Kolesnikov, Alexander and Weissenborn, Dirk and Zhai, Xiaohua and Unterthiner, Thomas and Dehghani, Mostafa and Minderer, Matthias and Heigold, Georg and Gelly, Sylvain and Uszkoreit, Jakob and Houlsby, Neil},
	month = jun,
	year = {2021},
	note = {arXiv:2010.11929 [cs]},
}

@misc{he_deep_2015,
	title = {Deep {Residual} {Learning} for {Image} {Recognition}},
	url = {http://arxiv.org/abs/1512.03385},
	doi = {10.48550/arXiv.1512.03385},
	urldate = {2024-06-17},
	publisher = {arXiv},
	author = {He, Kaiming and Zhang, Xiangyu and Ren, Shaoqing and Sun, Jian},
	month = dec,
	year = {2015},
	note = {arXiv:1512.03385 [cs]},
}

@misc{noauthor_imagenet_nodate,
	title = {{ImageNet}: {A} large-scale hierarchical image database {\textbar} {IEEE} {Conference} {Publication} {\textbar} {IEEE} {Xplore}},
	url = {https://ieeexplore.ieee.org/document/5206848},
	urldate = {2024-06-17},
}

@misc{dubey_computer-aided_2020,
	title = {Computer-aided abnormality detection in chest radiographs in a clinical setting via domain-adaptation},
	url = {http://arxiv.org/abs/2012.10564},
	doi = {10.48550/arXiv.2012.10564},
	urldate = {2024-06-20},
	publisher = {arXiv},
	author = {Dubey, Abhishek K. and Young, Michael T. and Stanley, Christopher and Lunga, Dalton and Hinkle, Jacob},
	month = dec,
	year = {2020},
	note = {arXiv:2012.10564 [cs]},
}

@misc{noauthor_deep-chest_nodate,
	title = {Deep-chest: {Multi}-classification deep learning model for diagnosing {COVID}-19, pneumonia, and lung cancer chest diseases - {ScienceDirect}},
	url = {https://www.sciencedirect.com/science/article/pii/S0010482521001426},
	urldate = {2024-06-20},
}

@article{singh_3d_2020,
	title = {{3D} {Deep} {Learning} on {Medical} {Images}: {A} {Review}},
	volume = {20},
	issn = {1424-8220},
	shorttitle = {{3D} {Deep} {Learning} on {Medical} {Images}},
	url = {https://www.ncbi.nlm.nih.gov/pmc/articles/PMC7570704/},
	doi = {10.3390/s20185097},
	number = {18},
	urldate = {2024-06-20},
	journal = {Sensors (Basel, Switzerland)},
	author = {Singh, Satya P. and Wang, Lipo and Gupta, Sukrit and Goli, Haveesh and Padmanabhan, Parasuraman and Gulyás, Balázs},
	month = sep,
	year = {2020},
	pmid = {32906819},
	pmcid = {PMC7570704},
	pages = {5097},
}

@misc{chen_generating_2022,
	title = {Generating {Radiology} {Reports} via {Memory}-driven {Transformer}},
	url = {http://arxiv.org/abs/2010.16056},
	doi = {10.48550/arXiv.2010.16056},
	urldate = {2024-06-23},
	publisher = {arXiv},
	author = {Chen, Zhihong and Song, Yan and Chang, Tsung-Hui and Wan, Xiang},
	month = apr,
	year = {2022},
	note = {arXiv:2010.16056 [cs]},
}

@inproceedings{jing_show_2019,
	title = {Show, {Describe} and {Conclude}: {On} {Exploiting} the {Structure} {Information} of {Chest} {X}-{Ray} {Reports}},
	shorttitle = {Show, {Describe} and {Conclude}},
	url = {http://arxiv.org/abs/2004.12274},
	doi = {10.18653/v1/P19-1657},
	urldate = {2024-06-23},
	booktitle = {Proceedings of the 57th {Annual} {Meeting} of the {Association} for {Computational} {Linguistics}},
	author = {Jing, Baoyu and Wang, Zeya and Xing, Eric},
	year = {2019},
	note = {arXiv:2004.12274 [cs, eess]},
	pages = {6570--6580},
}

@inproceedings{jing_automatic_2018,
	address = {Melbourne, Australia},
	title = {On the {Automatic} {Generation} of {Medical} {Imaging} {Reports}},
	url = {https://aclanthology.org/P18-1240},
	doi = {10.18653/v1/P18-1240},
	urldate = {2024-06-23},
	booktitle = {Proceedings of the 56th {Annual} {Meeting} of the {Association} for {Computational} {Linguistics} ({Volume} 1: {Long} {Papers})},
	publisher = {Association for Computational Linguistics},
	author = {Jing, Baoyu and Xie, Pengtao and Xing, Eric},
	editor = {Gurevych, Iryna and Miyao, Yusuke},
	month = jul,
	year = {2018},
	pages = {2577--2586},
}

@misc{noauthor_ybrid_nodate,
	title = {y?brid retrieval-generation reinforced agent for medical image report generation - {Recherche} {Google}},
	url = {https://www.google.com/search?q=y%02brid+retrieval-generation+reinforced+agent+for+medical+image+report+generation&rlz=1C1VDKB_frFR1102FR1102&oq=y%02brid+retrieval-generation+reinforced+agent+for+medical+image+report+generation&gs_lcrp=EgZjaHJvbWUyBggAEEUYOdIBBzMyNmowajeoAgCwAgA&sourceid=chrome&ie=UTF-8},
	urldate = {2024-06-23},
}

@misc{you_aligntransformer_2022,
	title = {{AlignTransformer}: {Hierarchical} {Alignment} of {Visual} {Regions} and {Disease} {Tags} for {Medical} {Report} {Generation}},
	shorttitle = {{AlignTransformer}},
	url = {http://arxiv.org/abs/2203.10095},
	doi = {10.48550/arXiv.2203.10095},
	urldate = {2024-06-23},
	publisher = {arXiv},
	author = {You, Di and Liu, Fenglin and Ge, Shen and Xie, Xiaoxia and Zhang, Jing and Wu, Xian},
	month = mar,
	year = {2022},
	note = {arXiv:2203.10095 [cs, eess]},
}

@misc{noauthor_large_nodate,
	title = {Large {Model} driven {Radiology} {Report} {Generation} with {Clinical} {Quality} {Reinforcement} {Learning}},
	url = {https://arxiv.org/html/2403.06728v1},
	urldate = {2024-06-23},
}

@misc{liu_clinically_2019,
	title = {Clinically {Accurate} {Chest} {X}-{Ray} {Report} {Generation}},
	url = {http://arxiv.org/abs/1904.02633},
	doi = {10.48550/arXiv.1904.02633},
	urldate = {2024-06-23},
	publisher = {arXiv},
	author = {Liu, Guanxiong and Hsu, Tzu-Ming Harry and McDermott, Matthew and Boag, Willie and Weng, Wei-Hung and Szolovits, Peter and Ghassemi, Marzyeh},
	month = jul,
	year = {2019},
	note = {arXiv:1904.02633 [cs]},
}

@misc{nooralahzadeh_progressive_2021,
	title = {Progressive {Transformer}-{Based} {Generation} of {Radiology} {Reports}},
	url = {http://arxiv.org/abs/2102.09777},
	doi = {10.48550/arXiv.2102.09777},
	urldate = {2024-06-23},
	publisher = {arXiv},
	author = {Nooralahzadeh, Farhad and Gonzalez, Nicolas Perez and Frauenfelder, Thomas and Fujimoto, Koji and Krauthammer, Michael},
	month = aug,
	year = {2021},
	note = {arXiv:2102.09777 [cs]},
}

@inproceedings{wang_inclusive_2022,
	address = {Cham},
	title = {An {Inclusive} {Task}-{Aware} {Framework} for {Radiology} {Report} {Generation}},
	isbn = {978-3-031-16452-1},
	doi = {10.1007/978-3-031-16452-1_54},
	language = {en},
	booktitle = {Medical {Image} {Computing} and {Computer} {Assisted} {Intervention} – {MICCAI} 2022},
	publisher = {Springer Nature Switzerland},
	author = {Wang, Lin and Ning, Munan and Lu, Donghuan and Wei, Dong and Zheng, Yefeng and Chen, Jie},
	editor = {Wang, Linwei and Dou, Qi and Fletcher, P. Thomas and Speidel, Stefanie and Li, Shuo},
	year = {2022},
	pages = {568--577},
}

@misc{papanikolaou_dare_2020,
	title = {{DARE}: {Data} {Augmented} {Relation} {Extraction} with {GPT}-2},
	shorttitle = {{DARE}},
	url = {http://arxiv.org/abs/2004.13845},
	doi = {10.48550/arXiv.2004.13845},
	urldate = {2024-06-24},
	publisher = {arXiv},
	author = {Papanikolaou, Yannis and Pierleoni, Andrea},
	month = apr,
	year = {2020},
	note = {arXiv:2004.13845 [cs, stat]},
}

@article{radford_language_nodate,
	title = {Language {Models} are {Unsupervised} {Multitask} {Learners}},
	language = {en},
	author = {Radford, Alec and Wu, Jeffrey and Child, Rewon and Luan, David and Amodei, Dario and Sutskever, Ilya},
}

@article{lavie_meteor_2009,
	title = {The {Meteor} metric for automatic evaluation of machine translation},
	volume = {23},
	issn = {1573-0573},
	url = {https://doi.org/10.1007/s10590-009-9059-4},
	doi = {10.1007/s10590-009-9059-4},
	language = {en},
	number = {2},
	urldate = {2024-06-24},
	journal = {Machine Translation},
	author = {Lavie, Alon and Denkowski, Michael J.},
	month = sep,
	year = {2009},
	pages = {105--115},
}

@book{noauthor_text_2004,
	address = {Barcelona, Spain},
	title = {Text {Summarization} {Branches} {Out}},
	url = {https://aclanthology.org/W04-1000},
	urldate = {2024-06-24},
	publisher = {Association for Computational Linguistics},
	month = jul,
	year = {2004},
}

@inproceedings{papineni_bleu_2002,
	address = {Philadelphia, Pennsylvania, USA},
	title = {Bleu: a {Method} for {Automatic} {Evaluation} of {Machine} {Translation}},
	shorttitle = {Bleu},
	url = {https://aclanthology.org/P02-1040},
	doi = {10.3115/1073083.1073135},
	urldate = {2024-06-24},
	booktitle = {Proceedings of the 40th {Annual} {Meeting} of the {Association} for {Computational} {Linguistics}},
	publisher = {Association for Computational Linguistics},
	author = {Papineni, Kishore and Roukos, Salim and Ward, Todd and Zhu, Wei-Jing},
	editor = {Isabelle, Pierre and Charniak, Eugene and Lin, Dekang},
	month = jul,
	year = {2002},
	pages = {311--318},
}

@misc{levy_same_2024,
	title = {Same {Task}, {More} {Tokens}: the {Impact} of {Input} {Length} on the {Reasoning} {Performance} of {Large} {Language} {Models}},
	shorttitle = {Same {Task}, {More} {Tokens}},
	url = {http://arxiv.org/abs/2402.14848},
	doi = {10.48550/arXiv.2402.14848},
	urldate = {2024-06-25},
	publisher = {arXiv},
	author = {Levy, Mosh and Jacoby, Alon and Goldberg, Yoav},
	month = feb,
	year = {2024},
	note = {arXiv:2402.14848 [cs]},
}

@misc{vaswani_attention_2023,
	title = {Attention {Is} {All} {You} {Need}},
	url = {http://arxiv.org/abs/1706.03762},
	doi = {10.48550/arXiv.1706.03762},
	urldate = {2024-06-25},
	publisher = {arXiv},
	author = {Vaswani, Ashish and Shazeer, Noam and Parmar, Niki and Uszkoreit, Jakob and Jones, Llion and Gomez, Aidan N. and Kaiser, Lukasz and Polosukhin, Illia},
	month = aug,
	year = {2023},
	note = {arXiv:1706.03762 [cs]},
}

@misc{powers_evaluation_2020,
	title = {Evaluation: from precision, recall and {F}-measure to {ROC}, informedness, markedness and correlation},
	shorttitle = {Evaluation},
	url = {http://arxiv.org/abs/2010.16061},
	doi = {10.48550/arXiv.2010.16061},
	urldate = {2024-06-26},
	publisher = {arXiv},
	author = {Powers, David M. W.},
	month = oct,
	year = {2020},
	note = {arXiv:2010.16061 [cs, stat]},
}

@article{sasaki_truth_nodate,
	title = {The truth of the {F}-measure},
	language = {en},
	author = {Sasaki, Yutaka},
}

@incollection{denotter_hounsfield_2024,
	address = {Treasure Island (FL)},
	title = {Hounsfield {Unit}},
	copyright = {Copyright © 2024, StatPearls Publishing LLC.},
	url = {http://www.ncbi.nlm.nih.gov/books/NBK547721/},
	language = {eng},
	urldate = {2024-06-27},
	booktitle = {{StatPearls}},
	publisher = {StatPearls Publishing},
	author = {DenOtter, Tami D. and Schubert, Johanna},
	year = {2024},
	pmid = {31613501},
}

@misc{chen_simple_2020,
	title = {A {Simple} {Framework} for {Contrastive} {Learning} of {Visual} {Representations}},
	url = {http://arxiv.org/abs/2002.05709},
	doi = {10.48550/arXiv.2002.05709},
	urldate = {2024-07-01},
	publisher = {arXiv},
	author = {Chen, Ting and Kornblith, Simon and Norouzi, Mohammad and Hinton, Geoffrey},
	month = jun,
	year = {2020},
	note = {arXiv:2002.05709 [cs, stat]},
}

@article{mai_systematic_2023,
	title = {A systematic review of automated segmentation of {3D} computed‐tomography scans for volumetric body composition analysis},
	volume = {14},
	issn = {2190-5991},
	url = {https://www.ncbi.nlm.nih.gov/pmc/articles/PMC10570079/},
	doi = {10.1002/jcsm.13310},
	number = {5},
	urldate = {2024-07-09},
	journal = {Journal of Cachexia, Sarcopenia and Muscle},
	author = {Mai, Dinh Van Chi and Drami, Ioanna and Pring, Edward T. and Gould, Laura E. and Lung, Phillip and Popuri, Karteek and Chow, Vincent and Beg, Mirza F. and Athanasiou, Thanos and Jenkins, John T.},
	month = aug,
	year = {2023},
	pmid = {37562946},
	pmcid = {PMC10570079},
	pages = {1973--1986},
}

@article{ilesanmi_reviewing_2024,
	title = {Reviewing {3D} convolutional neural network approaches for medical image segmentation},
	volume = {10},
	issn = {2405-8440},
	url = {https://www.sciencedirect.com/science/article/pii/S2405844024034297},
	doi = {10.1016/j.heliyon.2024.e27398},
	number = {6},
	urldate = {2024-07-09},
	journal = {Heliyon},
	author = {Ilesanmi, Ademola E. and Ilesanmi, Taiwo O. and Ajayi, Babatunde O.},
	month = mar,
	year = {2024},
	pages = {e27398},
}

@article{ranschaert_optimization_2021,
	title = {Optimization of {Radiology} {Workflow} with {Artificial} {Intelligence}},
	volume = {59},
	issn = {1557-8275},
	doi = {10.1016/j.rcl.2021.06.006},
	language = {eng},
	number = {6},
	journal = {Radiologic Clinics of North America},
	author = {Ranschaert, Erik and Topff, Laurens and Pianykh, Oleg},
	month = nov,
	year = {2021},
	pmid = {34689880},
	pages = {955--966},
}

@misc{oshea_introduction_2015,
	title = {An {Introduction} to {Convolutional} {Neural} {Networks}},
	url = {http://arxiv.org/abs/1511.08458},
	doi = {10.48550/arXiv.1511.08458},
	urldate = {2024-07-09},
	publisher = {arXiv},
	author = {O'Shea, Keiron and Nash, Ryan},
	month = dec,
	year = {2015},
	note = {arXiv:1511.08458 [cs]},
}

@article{sherstinsky_fundamentals_2020,
	title = {Fundamentals of {Recurrent} {Neural} {Network} ({RNN}) and {Long} {Short}-{Term} {Memory} ({LSTM}) {Network}},
	volume = {404},
	issn = {01672789},
	url = {http://arxiv.org/abs/1808.03314},
	doi = {10.1016/j.physd.2019.132306},
	urldate = {2024-07-09},
	journal = {Physica D: Nonlinear Phenomena},
	author = {Sherstinsky, Alex},
	month = mar,
	year = {2020},
	note = {arXiv:1808.03314 [cs, stat]},
	pages = {132306},
}

@article{bogatinovski_comprehensive_2022,
	title = {Comprehensive comparative study of multi-label classification methods},
	volume = {203},
	issn = {0957-4174},
	url = {https://www.sciencedirect.com/science/article/pii/S0957417422005991},
	doi = {10.1016/j.eswa.2022.117215},
	urldate = {2024-07-09},
	journal = {Expert Systems with Applications},
	author = {Bogatinovski, Jasmin and Todorovski, Ljupčo and Džeroski, Sašo and Kocev, Dragi},
	month = oct,
	year = {2022},
	pages = {117215},
}

@misc{crawshaw_multi-task_2020,
	title = {Multi-{Task} {Learning} with {Deep} {Neural} {Networks}: {A} {Survey}},
	shorttitle = {Multi-{Task} {Learning} with {Deep} {Neural} {Networks}},
	url = {http://arxiv.org/abs/2009.09796},
	doi = {10.48550/arXiv.2009.09796},
	urldate = {2024-07-09},
	publisher = {arXiv},
	author = {Crawshaw, Michael},
	month = sep,
	year = {2020},
	note = {arXiv:2009.09796 [cs, stat]},
}

@article{lopez-sanchez_supervised_2023,
	title = {Supervised {Deep} {Learning} {Techniques} for {Image} {Description}: {A} {Systematic} {Review}},
	volume = {25},
	issn = {1099-4300},
	shorttitle = {Supervised {Deep} {Learning} {Techniques} for {Image} {Description}},
	url = {https://www.ncbi.nlm.nih.gov/pmc/articles/PMC10138089/},
	doi = {10.3390/e25040553},
	number = {4},
	urldate = {2024-07-12},
	journal = {Entropy},
	author = {López-Sánchez, Marco and Hernández-Ocaña, Betania and Chávez-Bosquez, Oscar and Hernández-Torruco, José},
	month = mar,
	year = {2023},
	pmid = {37190341},
	pmcid = {PMC10138089},
	pages = {553},
}

@article{gaurav_survey_2021,
	title = {A {Survey} on {Various} {Deep} {Learning} {Models} for {Automatic} {Image} {Captioning}},
	volume = {1950},
	issn = {1742-6596},
	url = {https://dx.doi.org/10.1088/1742-6596/1950/1/012045},
	doi = {10.1088/1742-6596/1950/1/012045},
	language = {en},
	number = {1},
	urldate = {2024-07-12},
	journal = {Journal of Physics: Conference Series},
	author = {{Gaurav} and Mathur, Pratistha},
	month = aug,
	year = {2021},
	note = {Publisher: IOP Publishing},
	pages = {012045},
}

@article{wang_overview_2020,
	title = {An {Overview} of {Image} {Caption} {Generation} {Methods}},
	volume = {2020},
	issn = {1687-5265},
	url = {https://www.ncbi.nlm.nih.gov/pmc/articles/PMC7199544/},
	doi = {10.1155/2020/3062706},
	urldate = {2024-07-12},
	journal = {Computational Intelligence and Neuroscience},
	author = {Wang, Haoran and Zhang, Yue and Yu, Xiaosheng},
	month = jan,
	year = {2020},
	pmid = {32377178},
	pmcid = {PMC7199544},
	pages = {3062706},
}

@misc{chen_unit3d_2022,
	title = {{UniT3D}: {A} {Unified} {Transformer} for {3D} {Dense} {Captioning} and {Visual} {Grounding}},
	shorttitle = {{UniT3D}},
	url = {http://arxiv.org/abs/2212.00836},
	doi = {10.48550/arXiv.2212.00836},
	urldate = {2024-07-12},
	publisher = {arXiv},
	author = {Chen, Dave Zhenyu and Hu, Ronghang and Chen, Xinlei and Nießner, Matthias and Chang, Angel X.},
	month = dec,
	year = {2022},
	note = {arXiv:2212.00836 [cs]},
}

@misc{vinyals_show_2015,
	title = {Show and {Tell}: {A} {Neural} {Image} {Caption} {Generator}},
	shorttitle = {Show and {Tell}},
	url = {http://arxiv.org/abs/1411.4555},
	doi = {10.48550/arXiv.1411.4555},
	urldate = {2024-07-12},
	publisher = {arXiv},
	author = {Vinyals, Oriol and Toshev, Alexander and Bengio, Samy and Erhan, Dumitru},
	month = apr,
	year = {2015},
	note = {arXiv:1411.4555 [cs]},
}

@misc{zhang_remote_nodate,
	title = {Remote {Sensing} {\textbar} {Free} {Full}-{Text} {\textbar} {Description} {Generation} for {Remote} {Sensing} {Images} {Using} {Attribute} {Attention} {Mechanism}},
	url = {https://www.mdpi.com/2072-4292/11/6/612},
	urldate = {2024-07-12},
	author = {Zhang, Xiangrong and Wang, Xin and Tang, Xu and Zhou, Huiyu and Li, Chen},
}

@misc{gurari_vizwiz_2018,
	title = {{VizWiz} {Grand} {Challenge}: {Answering} {Visual} {Questions} from {Blind} {People}},
	shorttitle = {{VizWiz} {Grand} {Challenge}},
	url = {http://arxiv.org/abs/1802.08218},
	doi = {10.48550/arXiv.1802.08218},
	urldate = {2024-07-12},
	publisher = {arXiv},
	author = {Gurari, Danna and Li, Qing and Stangl, Abigale J. and Guo, Anhong and Lin, Chi and Grauman, Kristen and Luo, Jiebo and Bigham, Jeffrey P.},
	month = may,
	year = {2018},
	note = {arXiv:1802.08218 [cs]},
}

@misc{xue_multi-task_2024,
	title = {Multi-task {Prompt} {Words} {Learning} for {Social} {Media} {Content} {Generation}},
	url = {http://arxiv.org/abs/2407.07771},
	doi = {10.48550/arXiv.2407.07771},
	urldate = {2024-07-12},
	publisher = {arXiv},
	author = {Xue, Haochen and Zhang, Chong and Liu, Chengzhi and Wu, Fangyu and Jin, Xiaobo},
	month = jul,
	year = {2024},
	note = {arXiv:2407.07771 [cs]
version: 1},
}

@misc{kingma_adam_2017,
	title = {Adam: {A} {Method} for {Stochastic} {Optimization}},
	shorttitle = {Adam},
	url = {http://arxiv.org/abs/1412.6980},
	doi = {10.48550/arXiv.1412.6980},
	urldate = {2024-07-12},
	publisher = {arXiv},
	author = {Kingma, Diederik P. and Ba, Jimmy},
	month = jan,
	year = {2017},
	note = {arXiv:1412.6980 [cs]},
}

@misc{loshchilov_decoupled_2019,
	title = {Decoupled {Weight} {Decay} {Regularization}},
	url = {http://arxiv.org/abs/1711.05101},
	doi = {10.48550/arXiv.1711.05101},
	urldate = {2024-07-12},
	publisher = {arXiv},
	author = {Loshchilov, Ilya and Hutter, Frank},
	month = jan,
	year = {2019},
	note = {arXiv:1711.05101 [cs, math]},
}

@article{lecun_theoretical_2001,
	title = {A {Theoretical} {Framework} for {Back}-{Propagation}},
	author = {Lecun, Yann},
	month = aug,
	year = {2001},
}

@article{lowerre_harpy_nodate,
	title = {{THE} {HARPY} {SPEECH} {RECOGNITION} {SYSTEM}},
	language = {en},
	author = {Lowerre, Bruce T},
}

 \clearpage
\newpage
\onecolumn
\appendix
\section{Appendix: Qualitative results}
\label{sec:qualitative_results}

\begin{figure}[h!]
    \centering
    \vspace{-0.2cm}
    \begin{subfigure}[b]{0.9\linewidth}
        \includegraphics[width=\linewidth]{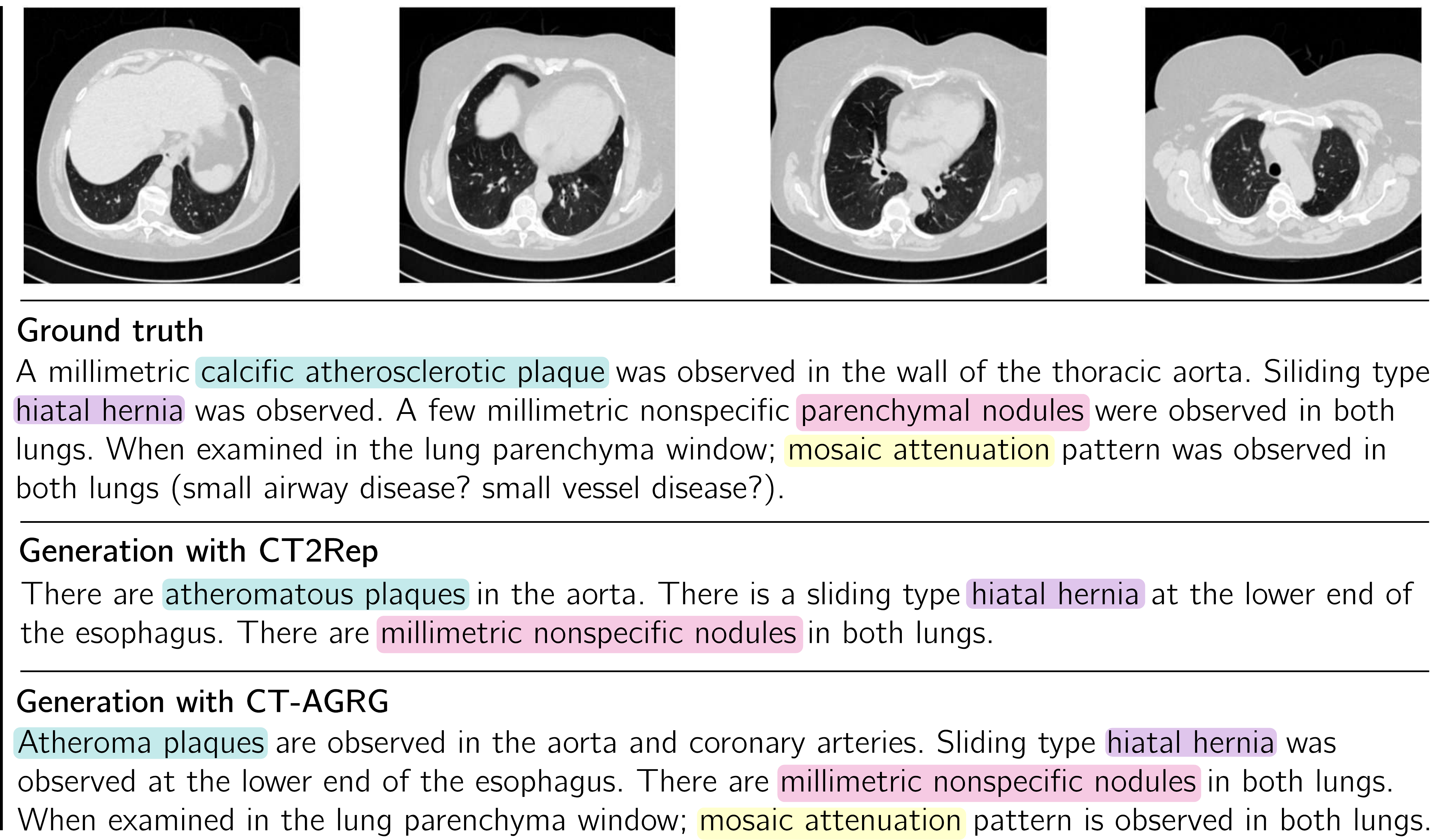}
        %\caption{\small Example 3}
        \label{fig:subfig3}
    \end{subfigure}
    \vspace{-0.2cm}
    \begin{subfigure}[b]{0.9\linewidth}
        \includegraphics[width=\linewidth]{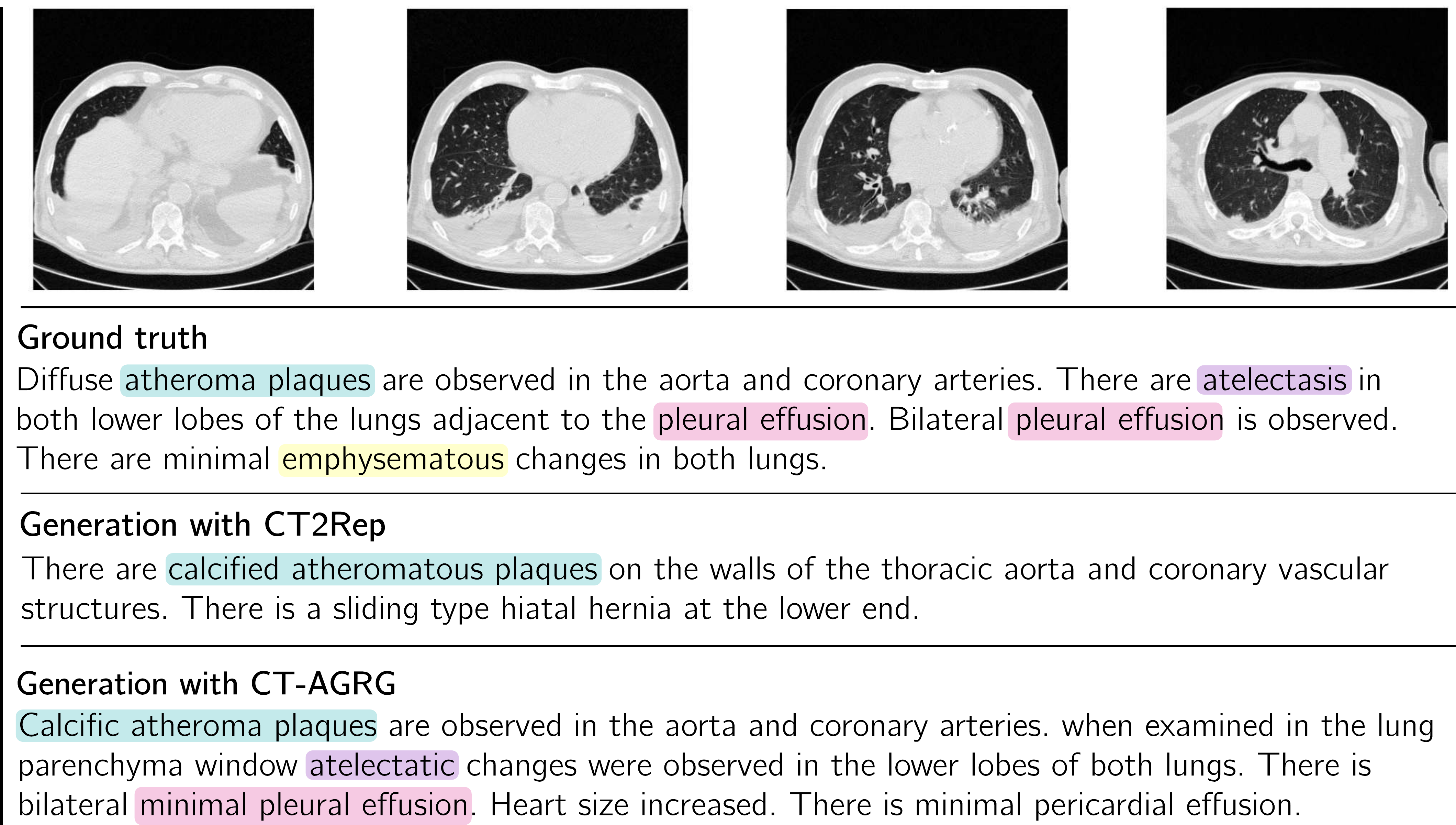}
        %\caption{\small Example 4}
        \label{fig:subfig4}
    \end{subfigure}
    \caption{Comparison of ground truth with reports generated by CT2Rep and CT-AGRG from the CT-RATE test set.}
    \label{fig:generations1}
\end{figure}

\begin{figure}[h!]
    \centering
    \begin{subfigure}[b]{0.8\linewidth}
        \includegraphics[width=\linewidth]{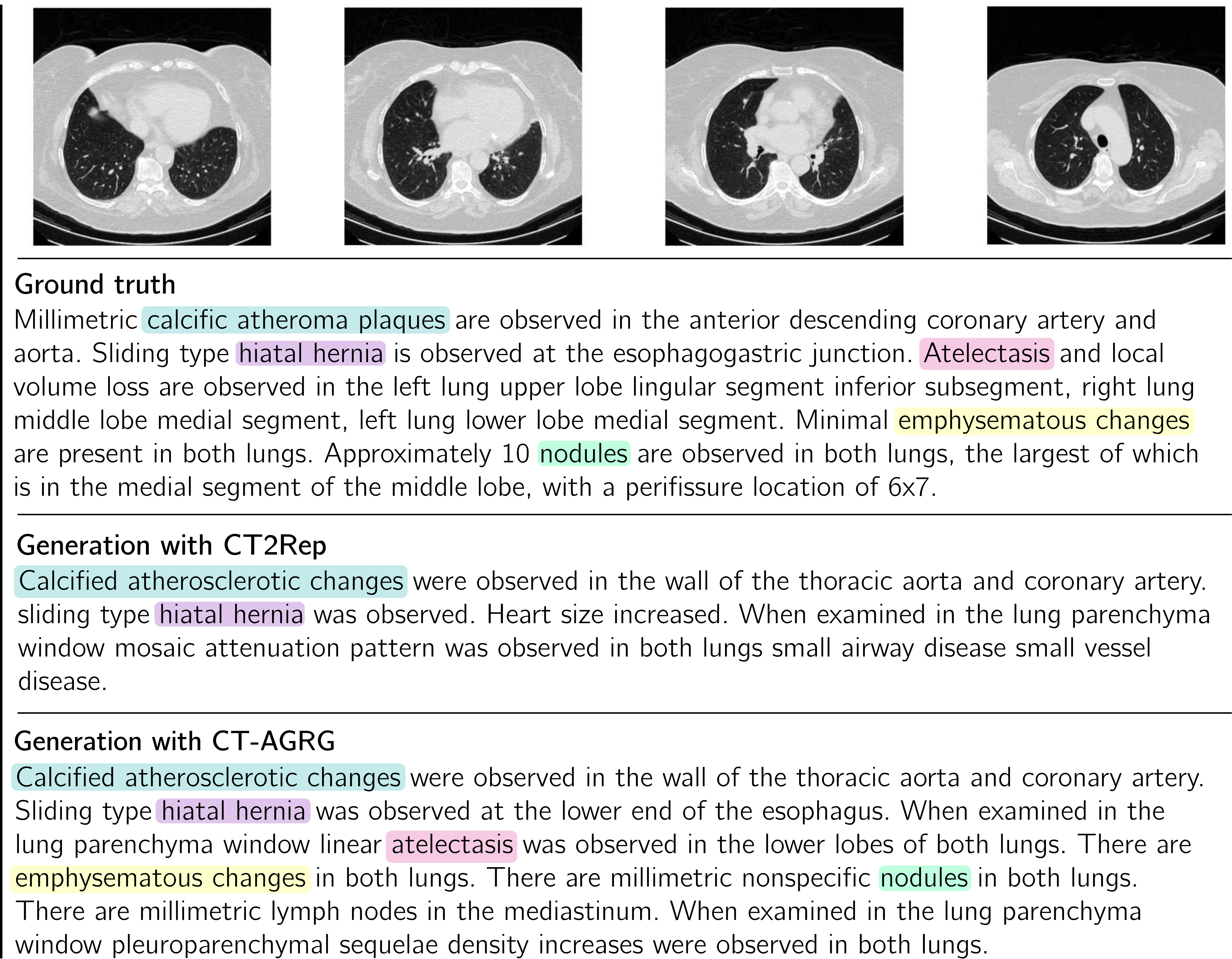}
        %\caption{\small Example 3}
        \label{fig:subfig3}
    \end{subfigure}
    \vspace{-0.2cm}
    \begin{subfigure}[b]{0.8\linewidth}
        \includegraphics[width=\linewidth]{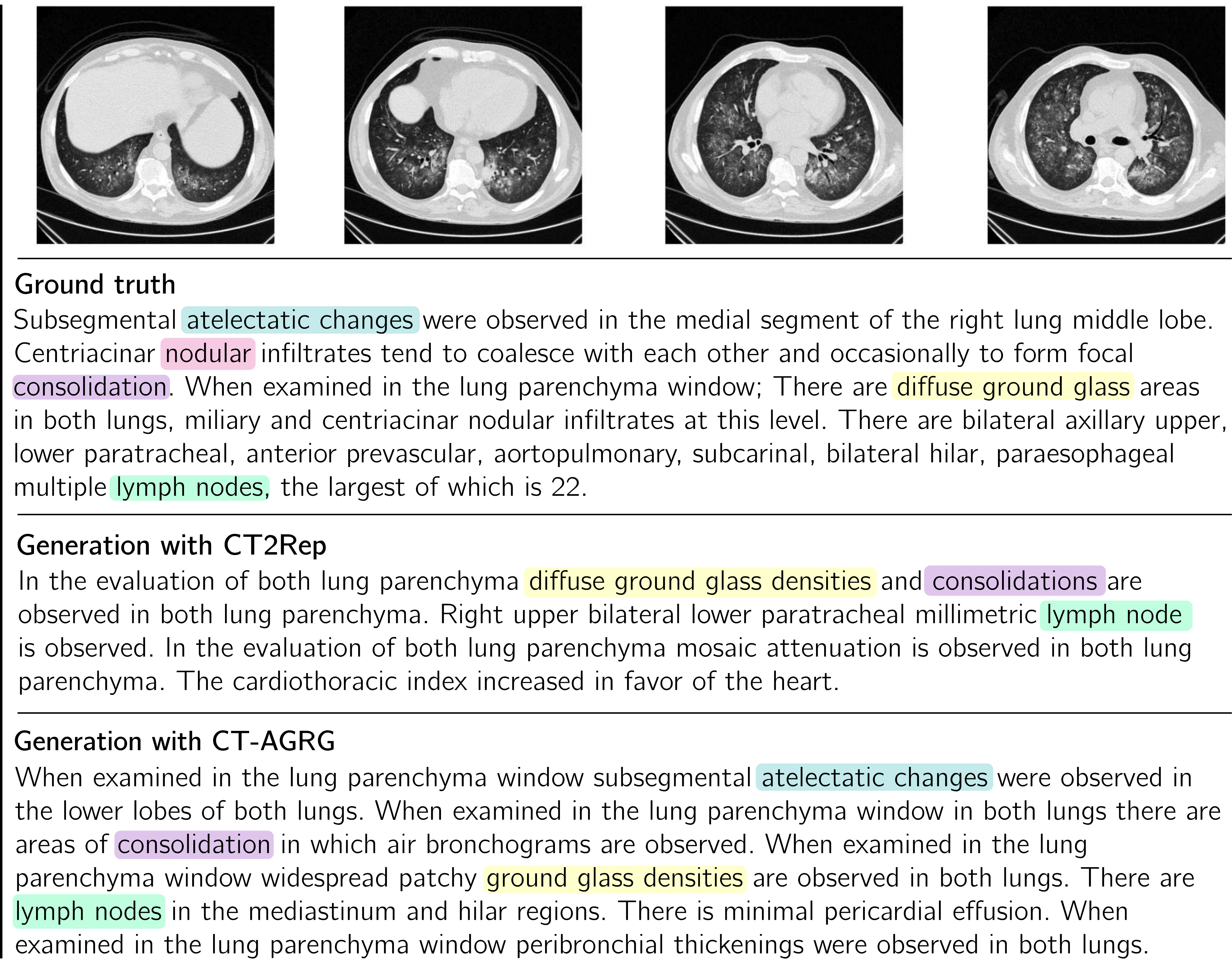}
        %\caption{\small Example 4}
        \label{fig:subfig4}
    \end{subfigure}
    \caption{Comparison of ground truth with reports generated by CT2Rep and CT-AGRG from the CT-RATE test set.}
    \label{fig:generations2}
\end{figure}

\end{document}